\documentclass[twocolumn,epjc3]{svjour3}
\usepackage[T1]{fontenc}
\usepackage[utf8]{inputenc}
\usepackage{graphicx,amsmath,amssymb,bm,mathptmx,microtype,ragged2e}
\usepackage{dblfloatfix}
\usepackage{flushend}
\usepackage[numbers,sort&compress]{natbib}
\usepackage[colorlinks=true,allcolors=blue]{hyperref}
\newcommand{\ReferenceLink}[2]{\href{#1}{\textcolor{blue}{#2}}}
\usepackage{refcount}
\newsavebox{\ExperimentalParameterTable}
\newsavebox{\ColumnEquationBox}
\newcommand{\FitColumnMath}[1]{%
  \begingroup
  \sbox{\ColumnEquationBox}{$\displaystyle #1$}%
  \ifdim\wd\ColumnEquationBox>\dimexpr\linewidth-\mathindent-2.5em\relax
    \resizebox{\dimexpr\linewidth-\mathindent-2.5em\relax}{!}{\usebox{\ColumnEquationBox}}%
  \else
    \usebox{\ColumnEquationBox}%
  \fi
  \endgroup
}

\DeclareRobustCommand{\hbar}{\mathord{\mathchar'26\mkern-9mu h}}
\journalname{Eur. Phys. J. C}
\smartqed
\AtBeginDocument{\raggedbottom\setlength{\parskip}{0pt}}
\makeatletter
\g@addto@macro\normalsize{%
  \setlength{\abovedisplayskip}{5pt plus 1pt minus 1pt}%
  \setlength{\belowdisplayskip}{5pt plus 1pt minus 1pt}%
  \setlength{\abovedisplayshortskip}{3pt plus 1pt minus 1pt}%
  \setlength{\belowdisplayshortskip}{3pt plus 1pt minus 1pt}}
\renewcommand\section{\@startsection{section}{1}{\z@}%
  {-12pt plus -2pt minus -2pt}{6pt}{\normalsize\bfseries\boldmath}}
\renewcommand\subsection{\@startsection{subsection}{2}{\z@}%
  {-9pt plus -2pt minus -1pt}{4pt}{\normalsize\upshape}}
\renewcommand\paragraph{\@startsection{paragraph}{4}{\z@}%
  {-6pt plus -1pt minus -1pt}{-.5em}{\normalsize\itshape}}
\newcount\FiguresOnCurrentPage
\newcommand{\CheckCaptionPage}[2]{%
  \@ifundefined{r@#1}{}{%
    \typeout{FloatLayout: #1 page \getpagerefnumber{#1}; target #2}%
    \ifnum\getpagerefnumber{#1}=#2\relax
    \else
      \PackageWarningNoLine{FloatLayout}{#1 is on page \getpagerefnumber{#1}, target #2.
        This is an advisory layout notice, not a compilation error}%
    \fi
  }%
}
\AtEndDocument{%
  \CheckCaptionPage{tab:experiments}{3}%
  \CheckCaptionPage{fig:3}{6}%
  \CheckCaptionPage{fig:4}{7}%
  \begingroup
    \@tempcnta=0\relax
    \FiguresOnCurrentPage=0\relax
    \@for\FigureIndex:=1,2,3,4,5\do{%
      \@ifundefined{r@fig:\FigureIndex}{}{%
        \@tempcntb=\getpagerefnumber{fig:\FigureIndex}\relax
        \typeout{FloatLayout: figure \FigureIndex\space page \the\@tempcntb}%
        \ifnum\@tempcntb=\@tempcnta
          \advance\FiguresOnCurrentPage by 1
        \else
          \FiguresOnCurrentPage=1\relax
        \fi
        \ifnum\FiguresOnCurrentPage>2
          \PackageWarningNoLine{FloatLayout}{More than two figures on page \the\@tempcntb}%
        \fi
        \@tempcnta=\@tempcntb
      }%
    }%
  \endgroup
}
\makeatother
\title{Unifying First-Order Coherence, Concurrence, and Bell Correlations in Three-Flavor Neutrino Oscillations}
\titlerunning{Unified coherence, concurrence and Bell relations in neutrino oscillations}
\author{Bingbing Li\thanksref{aff:li}
\and Tao Wu\thanksref{aff:wu}
\and Jiadong Shi\thanksref{aff:shi,correspondingshi}}
\thankstext[*]{correspondingshi}{\email{\href{mailto:ah_sjd@126.com}{ah\_sjd@126.com}} (corresponding author)}
\authorrunning{Bingbing Li et al.}
\institute{Faculty of Engineering, Anhui Sanlian University, Hefei 230601, China\label{aff:li}
\and School of Physics \& Electronics Engineering, Fuyang Normal University, Fuyang 236037, China\label{aff:wu}
\and School of Physics \& Electronics Engineering, Hefei Institute of Technology, Hefei 238706, China\label{aff:shi}}
\date{}
\begin{document}
\maketitle
\begin{abstract}
Neutrino oscillations provide a natural setting for investigating the quantum character of flavor conversion and the distribution of quantum resources over macroscopic distances. Here, we study exact relations among purity-based first-order coherence, pairwise concurrence, and the maximal Clauser--Horne--Shimony--Holt (CHSH) value in coherent three-flavor neutrino oscillations. By expressing the evolving neutrino as a three-qubit single-excitation state, we derive the generally mixed two-qubit reductions and show that the squared first-order coherence of each flavor mode and its two associated squared concurrences sum to unity. Furthermore, a closed two-branch relation is obtained, showing that the maximal CHSH value of any reduced pair is determined by its average first-order coherence and concurrence. The branch structure follows from the correlation-tensor spectrum and yields exact conditions for CHSH violation. For the vacuum parameters and propagation ranges considered, electron preparation gives weak CHSH violations in electron-containing pairs, while muon preparation produces a dominant \(\mu\tau\) pair with substantially stronger concurrence and Bell correlations. The comparison also identifies the third-flavor contribution responsible for the different locations of concurrence and CHSH maxima. These results clarify the distribution of quantum correlations during flavor evolution and provide a basis for extensions to coherent matter propagation and wave-packet-induced mixed states.
\end{abstract}
\section{Introduction}
Neutrino oscillations arise from coherent mixing of mass eigenstates and describe flavor conversion during propagation \cite{maki1962,giuntikimbook,akhmedov2009}. Atmospheric and solar observations established flavor transformation \cite{superk1998,sno2002}, while reactor and accelerator disappearance measurements probe complementary oscillation scales \cite{dayabay2012,minos2008,minos2014,dayabay2015}. In the occupation-number representation, the electron, muon, and tau modes form a three-qubit system restricted to a single excitation, allowing flavor conversion to be studied through single-particle mode entanglement \cite{blasone2008mixing,blasone2008}.

Quantum-information studies of neutrino oscillations have examined entanglement and Bell correlations in three-flavor mixing \cite{banerjee2015three,banerjee2015}, coherence in observed oscillations \cite{song2018}, temporal correlations \cite{formaggio2016}, and comparisons of several correlation quantifiers \cite{ming2020}. Tradeoffs involving first-order coherence, intrinsic concurrence, and pairwise CHSH tests \cite{li2022}, together with complete complementarity analyses \cite{bittencourt2022,bittencourt2024}, establish a basis for studying resource distribution among flavor modes. Wave-packet treatments and experimental coherence studies address the propagation assumptions \cite{blasone2021,dayabay2017}, while general three-qubit results connect coherence with entanglement and reduced-pair correlations \cite{dong2022,dong2023}. For the single-excitation state, tracing out an occupied flavor leaves a vacuum contribution in the selected pair, which must be retained when relating concurrence to the maximal CHSH value.

Quantum resource theories quantify state properties under specified restrictions \cite{chitambar2019,horodecki2009,streltsov2017}. Basis-dependent coherence \cite{baumgratz2014}, purity-based first-order coherence \cite{svozilik2015}, CHSH correlations \cite{chsh1969,horodecki1995}, and steering \cite{wiseman2007,uola2020,cavalcanti2009,costa2016} have distinct definitions and operational roles, including connections with teleportation \cite{fan2022}. Complementarity constrains local properties and correlations through wave--particle duality and entanglement-completed balances \cite{englert1996,jakob2010}, with complete formulations for multipartite pure and mixed states \cite{basso2020,bassomixed2021}. Related coherence--steering and entanglement tradeoffs \cite{dutong2021,ming2021}, as well as unified interferometric relations \cite{gravity2024}, motivate deriving exact identities for the particular reduced states generated by neutrino mixing.

In this paper, we derive the first-order coherence--concurrence balance for each reference flavor and a closed relation among average first-order coherence, ordinary concurrence, and the optimized CHSH value of any reduced pair. The derivation uses the purity of the complete state and the spectrum of the pair correlation tensor; the pair itself is generally mixed. We then examine the relations for initially electron and muon neutrinos. The comparison shows how the third flavor changes Bell-violating intervals and the locations of concurrence and CHSH maxima. Experimental disappearance points accompany the survival curves, while the correlation curves use the complete three-flavor probabilities: one survival probability alone does not determine the two separate appearance channels.

The paper is organized as follows. Section 2 introduces three-flavor propagation and the reduced two-qubit states. Section 3 defines the three quantifiers and derives their exact relations. Section 4 examines the flavor-dependent evolution for electron and muon preparation. Section 5 presents the conclusions.

\section{Three-flavor neutrino oscillations}
\subsection{Flavor mixing and coherent propagation}
The three active neutrino flavors, \(\nu_e\), \(\nu_\mu\), and \(\nu_\tau\), are coherent superpositions of the mass eigenstates \(\nu_1\), \(\nu_2\), and \(\nu_3\). Following the flavor convention used in Refs.~\cite{giuntikimbook,li2022}, their relation and the standard mixing-matrix decomposition are
\begin{equation}\begin{gathered}
|\nu_\alpha\rangle=\sum_{r=1}^{3}U_{\alpha r}^{*}|\nu_r\rangle,
\qquad U=R_{23}U_{13}R_{12},\\
U_{13}=\begin{pmatrix}
c_{13}&0&s_{13}e^{-i\delta}\\
0&1&0\\
-s_{13}e^{i\delta}&0&c_{13}
\end{pmatrix}.
\end{gathered}\tag{1}\label{eq:1}\end{equation}
Here \(\alpha=e,\mu,\tau\), \(r=1,2,3\), and \(U\) is the unitary Pontecorvo--Maki--Nakagawa--Sakata (PMNS) matrix. We write \(s_{ab}=\sin\theta_{ab}\), \(c_{ab}=\cos\theta_{ab}\), and \(\delta=\delta_{\mathrm{CP}}\); \(R_{12}\) and \(R_{23}\) are real rotations in the indicated planes. Possible Majorana phases cancel from oscillation probabilities.

We use natural units, \(\hbar=c=1\). For an ultrarelativistic neutrino, \(E_r\simeq E+m_r^2/(2E)\) and the propagation distance is \(L\simeq t\). Removing the common kinematic phase, an initially prepared flavor evolves as
\begin{equation}
\FitColumnMath{|\nu_\alpha(L)\rangle=\sum_r U_{\alpha r}^{*}e^{-im_r^2L/(2E)}|\nu_r\rangle=\sum_\beta a_{\alpha\beta}(L)|\nu_\beta\rangle.}
\tag{2}\label{eq:2}
\end{equation}
Thus different propagation phases produce flavor conversion. The transition amplitude and probability can be written as
\begin{equation}\begin{gathered}
a_{\alpha\beta}(L)=\sum_r U_{\beta r}U_{\alpha r}^{*}e^{-im_r^2L/(2E)},\\
\FitColumnMath{P_{\alpha\beta}=\delta_{\alpha\beta}-4\sum_{r>s}\operatorname{Re}J_{\alpha\beta}^{rs}\sin^2\Delta_{rs}+2\sum_{r>s}\operatorname{Im}J_{\alpha\beta}^{rs}\sin(2\Delta_{rs}),}\\
J_{\alpha\beta}^{rs}=U_{\beta r}U_{\alpha r}^{*}U_{\beta s}^{*}U_{\alpha s}.
\end{gathered}\tag{3}\label{eq:3}\end{equation}
The oscillation phase is \(\Delta_{rs}=\Delta m_{rs}^2L/(4E)\), where \(\Delta m_{rs}^2=m_r^2-m_s^2\). In numerical calculations, \(L/E\) is in \(\mathrm{km/GeV}\) and the mass-squared splittings are in \(\mathrm{eV}^2\), giving \(\Delta_{rs}=1.266932679\,\Delta m_{rs}^2(L/E)\). For \(\beta=\alpha\), the imaginary term vanishes and Eq.~\eqref{eq:3} gives the survival probability. Unitarity ensures \(\sum_\beta P_{\alpha\beta}=1\); both independent mass splittings are retained.

This description assumes coherence between the propagating mass components. Wave-packet analyses specify the production, localization, and detection conditions for this approximation \cite{kayser1981,giuntikim1998,giunti2004,kiers1996,stodolsky1998}. Field-theoretical treatments place these stages in a common framework \cite{blasone1995,beuthe2003,akhmedovkopp2010}. We consider coherent vacuum propagation, for which the complete flavor state remains pure.

\begin{table*}[t]
\caption{Representative experimental scales and central oscillation parameters. Mass-squared splittings are in \(\mathrm{eV}^2\). The entries use the KamLAND-only three-flavor fit, the physical MINOS two-flavor fit, and the normal-ordering T2K neutrino-mode fit.}
\label{tab:experiments}
\centering
\small
\setlength{\tabcolsep}{4pt}
\renewcommand{\arraystretch}{1.18}
\sbox{\ExperimentalParameterTable}{%
\begin{tabular}{llclll}
\hline
Experiment & Beam & \(L\) (km) & Energy & Mass-squared splitting & Mixing parameter \\
\hline
Daya Bay \cite{dayabay2015} & \(\bar\nu_e\) & \(0.364\)--\(1.912\) & \(1\)--\(8\) MeV & \(|\Delta m_{ee}^2|=2.42\times10^{-3}\) & \(\sin^2(2\theta_{13})=0.084\) \\
KamLAND \cite{kamland2013} & \(\bar\nu_e\) & \(\simeq180\) & \(2\)--\(10\) MeV & \(\Delta m_{21}^2=7.54\times10^{-5}\) & \(\tan^2\theta_{12}=0.481\) \\
MINOS \cite{minos2011} & \(\nu_\mu\) & \(\simeq735\) & \(0.5\)--\(50\) GeV & \(|\Delta m_{\mathrm{eff}}^2|=2.32\times10^{-3}\) & \(\sin^2(2\theta_{\mathrm{eff}})=1.00\) \\
T2K \cite{t2k2017} & \(\nu_\mu\) & \(295\) & peak \(0.6\) GeV & \(\Delta m_{32}^2=2.53\times10^{-3}\) & \(\sin^2\theta_{23}=0.51\) \\
\hline
\end{tabular}
}
\ifdim\wd\ExperimentalParameterTable>\textwidth
  \PackageError{ExperimentalParameterTable}{Parameter table exceeds text width}{Reduce table padding or shorten the headings.}
\fi
\usebox{\ExperimentalParameterTable}
\end{table*}

\subsection{Three-qubit representation and reduced two-qubit states}
To study quantum correlations, we assign a qubit to each flavor mode: \(|1\rangle_i\) and \(|0\rangle_i\) denote occupation and absence of occupation, respectively. The flavor states then correspond to \(|\nu_e\rangle=|100\rangle\), \(|\nu_\mu\rangle=|010\rangle\), and \(|\nu_\tau\rangle=|001\rangle\) \cite{blasone2008,li2022}. Equation~\eqref{eq:3} gives the three-qubit state
\begin{equation}\begin{gathered}
|\psi_\alpha(L)\rangle=a_{\alpha e}|100\rangle
+a_{\alpha\mu}|010\rangle+a_{\alpha\tau}|001\rangle,\\
\rho_\alpha=|\psi_\alpha\rangle\langle\psi_\alpha|,\qquad
p_i=|a_{\alpha i}|^2=P_{\alpha i},\\
p_e+p_\mu+p_\tau=1.
\end{gathered}\tag{4}\label{eq:4}\end{equation}
The first index of \(P_{\alpha i}\) denotes the preparation and the second the detected flavor. The qubits represent three modes of one particle. We suppress the initial-flavor index in the correlation formulas below and use \(i,j,k\) for three distinct flavor modes, not mass eigenstates.

The reduced state of modes \(i,j\) follows by tracing out mode \(k\). The component occupying \(k\) contributes \(|00\rangle_{ij}\), whereas the components occupying \(i\) and \(j\) retain their mutual coherence. Explicitly,
\begin{equation}\begin{gathered}
\rho_{ij}=\operatorname{Tr}_k\rho_\alpha
=p_k|00\rangle\langle00|+|\chi_{ij}\rangle\langle\chi_{ij}|,\\
|\chi_{ij}\rangle=a_{\alpha i}|10\rangle+a_{\alpha j}|01\rangle,\\
\rho_{ij}=\begin{pmatrix}
p_k&0&0&0\\
0&p_j&z_{ij}^{*}&0\\
0&z_{ij}&p_i&0\\
0&0&0&0
\end{pmatrix},\\
z_{ij}=a_{\alpha i}a_{\alpha j}^{*},\qquad |z_{ij}|^2=p_ip_j.
\end{gathered}\tag{5}\label{eq:5}\end{equation}
The matrix basis is \(\{|00\rangle,|01\rangle,|10\rangle,|11\rangle\}\), and \(|\chi_{ij}\rangle\) has squared norm \(1-p_k\). Choosing \((i,j,k)=(e,\mu,\tau)\), \((e,\tau,\mu)\), and \((\mu,\tau,e)\) gives \(\rho_{e\mu}\), \(\rho_{e\tau}\), and \(\rho_{\mu\tau}\), respectively. Each reduced spectrum is \(\{p_k,1-p_k,0,0\}\). Thus a pure three-qubit flavor state generally produces mixed two-qubit reductions; the discarded-flavor population cannot be omitted. Further partial traces give
\begin{equation}\begin{aligned}
\rho_i&=\operatorname{diag}(1-p_i,p_i),\\
\rho_j&=\operatorname{diag}(1-p_j,p_j).
\end{aligned}\tag{6}\label{eq:6}\end{equation}
These reduced states provide the common starting point for the first-order coherence, concurrence, and Bell calculations in Sect.~3.

\subsection{Oscillation parameters}
\begin{figure*}[t]
\centering
\includegraphics[width=.90\textwidth]{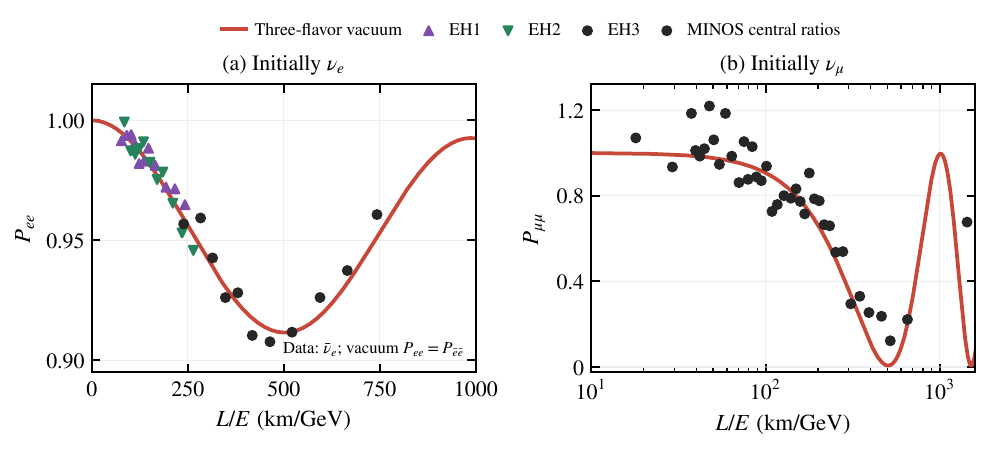}
\caption{Survival probabilities for the two initial flavors. Solid red curves are the three-flavor vacuum predictions from Eqs.~\eqref{eq:3} and~\eqref{eq:7}. (a) Electron-flavor survival with 33 reconstructed Daya Bay central markers from EH1, EH2, and EH3 \cite{dayabay2015,ming2020}. These are electron-antineutrino disappearance ratios; vacuum propagation gives \(P_{ee}=P_{\bar e\bar e}\). (b) Muon-flavor survival with 39 reconstructed MINOS disappearance/rate ratios \cite{sousa2015,ming2020}, retaining values above unity. The abscissa is linear in (a) and logarithmic in (b), in \(\mathrm{km/GeV}\). The unsmeared curves and central-point overlays illustrate the disappearance patterns.}
\label{fig:1}
\end{figure*}

The calculations use normal mass ordering and the fixed mixing angles and mass-squared differences \cite{nufit2024}
\begin{equation}\begin{gathered}\sin^2\theta_{12}=0.308,\quad\sin^2\theta_{13}=0.02215,\\\sin^2\theta_{23}=0.470,\quad\delta_{\mathrm{CP}}=212^\circ,\\\Delta m_{21}^2=7.49\times10^{-5}\,\mathrm{eV}^2,\\\Delta m_{31}^2=2.513\times10^{-3}\,\mathrm{eV}^2.\end{gathered}\tag{7}\label{eq:7}\end{equation}
\noindent The electron- and muon-initial curves cover \(0\le L/E\le1000\) and \(0\le L/E\le1600\,\mathrm{km/GeV}\), respectively. Table~\ref{tab:experiments} summarizes the experimental scales and central values for Daya Bay, KamLAND, MINOS, and T2K. Figure~\ref{fig:1} overlays reconstructed Daya Bay electron-antineutrino and MINOS muon-neutrino central ratios on the vacuum survival curves \cite{dayabay2015,sousa2015,ming2020}; vacuum electron-flavor disappearance satisfies \(P_{ee}=P_{\bar e\bar e}\). All correlation curves use the full normalized three-flavor probabilities.\label{discussion:survival}

\section{Unified relations among first-order coherence, concurrence, and Bell correlations}
\subsection{First-order coherence and concurrence}
Following the purity-based convention used in Refs. \cite{svozilik2015,dong2022,dong2023}, define

\begin{equation}\begin{aligned}D_i&=\sqrt{2\operatorname{Tr}\rho_i^2-1},\\D_{ij}^2&=\frac{D_i^2+D_j^2}{2},\\D^2&=\frac{D_e^2+D_\mu^2+D_\tau^2}{3}.\end{aligned}\tag{8}\label{eq:8}\end{equation}

All three quantities are dimensionless. The single-mode quantity \(D_i\) is the local Bloch-vector length; \(D_{ij}\) is its root-mean-square over the selected pair, whereas \(D\) averages all three modes. They must be distinguished when comparing different relations. Substitution of Eq.~\eqref{eq:6} gives

\begin{equation}\begin{aligned}D_i^2&=(1-2p_i)^2,\\D_{ij}^2&=\frac{(1-2p_i)^2+(1-2p_j)^2}{2}.\end{aligned}\tag{9}\label{eq:9}\end{equation}

This first-order coherence is invariant under local basis rotations and differs from occupation-basis off-diagonal coherence. In fact, the states in Eq.~\eqref{eq:6} are diagonal in that basis. Thus, a value \(D_i=1\) indicates a pure local occupation state, whereas \(D_i=0\) indicates a maximally mixed local state. We retain the established first-order-coherence terminology while using its explicit purity definition throughout. Resource conversions between basis coherence and entanglement \cite{streltsov2015}, and conditional protocols underlying the nonlocal advantage of coherence \cite{mondal2017}, concern different constructions and do not identify those quantities with \(D_i\).

\paragraph{Pairwise concurrence.}

For a two-qubit state, Wootters concurrence is \cite{wootters1998}

\begin{equation}\begin{aligned}C_{ij}&=\max\{0,\lambda_1-\lambda_2-\lambda_3-\lambda_4\},\\\widetilde\rho_{ij}&=(\sigma_y\otimes\sigma_y)\rho_{ij}^*(\sigma_y\otimes\sigma_y),\end{aligned}\tag{10}\label{eq:10}\end{equation}
\noindent where \(\lambda_r\) are the nonnegative square roots of the eigenvalues of \(\rho_{ij}\widetilde\rho_{ij}\), in decreasing order. For Eq.~\eqref{eq:5}, these roots are \(2\sqrt{p_ip_j},0,0,0\). Equivalently, the X-state formula gives \(C_{ij}=2|z_{ij}|\), since the \(|11\rangle\) population vanishes. Therefore

\begin{equation}C_{ij}=2\sqrt{p_ip_j},\qquad C_{ij}^2=4p_ip_j.\tag{11}\label{eq:11}\end{equation}
\noindent Here \(C_{ij}\) measures entanglement of the two-mode reduced state and \(C_{ij}^2\) is its bipartite tangle. It is distinct from genuine tripartite concurrence. For the complete pure state, the one-versus-two concurrence is

\begin{equation}\FitColumnMath{C_{i|jk}^2=2\left[1-\operatorname{Tr}\rho_i^2\right]=4p_i(1-p_i).}\tag{12}\label{eq:12}\end{equation}

Equations~\eqref{eq:11} and~\eqref{eq:12} saturate the Coffman-Kundu-Wootters relation, \(C_{i|jk}^2=C_{ij}^2+C_{ik}^2\) \cite{coffman2000}. The residual three-tangle is zero, as expected for a single-excitation W-type state. This does not exclude genuine tripartite mode entanglement when all three amplitudes are nonzero.

\subsection{Bell--CHSH inequality and correlation tensor}

Let \(A_0,A_1\) and \(B_0,B_1\) be local dichotomic observables acting on modes \(i\) and \(j\). The CHSH operator is \cite{chsh1969}

\begin{equation}\FitColumnMath{\mathcal B_{ij}=A_0\otimes(B_0+B_1)+A_1\otimes(B_0-B_1).}\tag{13}\label{eq:13}\end{equation}

Local hidden-variable models satisfy \(|\langle\mathcal B_{ij}\rangle|\le2\). Define the optimized value

\begin{equation}\FitColumnMath{B_{ij}=\max_{A_0,A_1,B_0,B_1}|\operatorname{Tr}(\rho_{ij}\mathcal B_{ij})|=2\sqrt{u_1+u_2},}\tag{14}\label{eq:14}\end{equation}
\noindent where \(u_1,u_2\) are the two largest eigenvalues of \(T_{ij}^{\mathsf T}T_{ij}\). The correlation-tensor entries are

\begin{equation}\FitColumnMath{(T_{ij})_{mn}=\operatorname{Tr}[\rho_{ij}(\sigma_m\otimes\sigma_n)].}\tag{15}\label{eq:15}\end{equation}
with \(m,n=x,y,z\).

The last equality in Eq.~\eqref{eq:14} is the Horodecki criterion \cite{horodecki1995}. The raw optimized value satisfies \(0\le B_{ij}\le2\sqrt2\); violation occurs only for \(B_{ij}>2\). It is not itself an entanglement monotone. Local phase rotations can make \(z_{ij}\) real and positive without changing \(D_i\), \(C_{ij}\), or \(B_{ij}\). In that frame,

\begin{equation}\begin{aligned}T_{ij}&=\operatorname{diag}(C_{ij},C_{ij},2p_k-1),\\\operatorname{spec}(T_{ij}^{\mathsf T}T_{ij})&=\{C_{ij}^2,C_{ij}^2,D_k^2\}.\end{aligned}\tag{16}\label{eq:16}\end{equation}
\noindent Thus the exact two-mode Bell expression is

\begin{equation}B_{ij}=2\sqrt{C_{ij}^2+\max\{C_{ij}^2,D_k^2\}}.\tag{17}\label{eq:17}\end{equation}
\noindent The third flavor enters through \(D_k^2=(1-2p_k)^2\). Consequently, concurrence alone does not fix the maximal CHSH value of a general reduced pair.

\begin{figure*}[t]\centering\includegraphics[width=\textwidth]{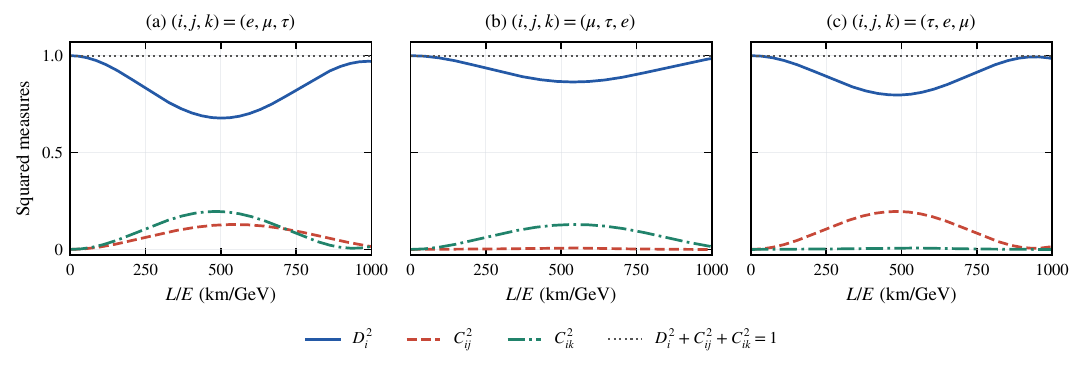}\caption{First-order coherence and pairwise concurrence for initially electron neutrinos. Panels (a)--(c) use the cyclic mode orderings \((i,j,k)=(e,\mu,\tau)\), \((\mu,\tau,e)\), and \((\tau,e,\mu)\). Solid blue, dashed red, and dash-dotted green curves show \(D_i^2\), \(C_{ij}^2\), and \(C_{ik}^2\), respectively, versus \(L/E\). The dotted curve is their sum, which remains unity by Eq.~\eqref{eq:18}. All curves are theoretical predictions.}\label{fig:2}\end{figure*}

\subsection{First-order coherence--concurrence balances}

\begin{corollary}\label{cor:local}
For a normalized pure state in the three-mode single-excitation sector, the first-order coherence of mode \(i\) and its two incident pairwise concurrences satisfy
\begin{equation}D_i^2+C_{ij}^2+C_{ik}^2=1.\tag{18}\label{eq:18}\end{equation}
\end{corollary}

\begin{proof}
The first-order coherence and the one-versus-two concurrence of a pure state satisfy \(D_i^2+C_{i|jk}^2=1\). For the state in Eq.~\eqref{eq:4}, the normalization \(p_j+p_k=1-p_i\) and Eqs.~\eqref{eq:9} and~\eqref{eq:11} give
\begin{equation}\begin{aligned}D_i^2+C_{ij}^2+C_{ik}^2
&=(1-2p_i)^2+4p_i(p_j+p_k)\\
&=1-4p_i+4p_i^2+4p_i(1-p_i)\\
&=1.\end{aligned}\tag{19}\label{eq:19}\end{equation}
Equivalently, Eq.~\eqref{eq:12} saturates the monogamy relation \cite{coffman2000}. Combining it with the pure-state first-order coherence balance gives Eq.~\eqref{eq:18}, completing the proof.
\end{proof}

Equation~\eqref{eq:18} holds for each choice of the reference flavor. In the two-flavor restriction \(p_k=0\), \(C_{ik}\) vanishes and the relation reduces to \(D_i^2+C_{ij}^2=1\). For three flavors the contribution absent from a two-quantity comparison is exactly
\begin{equation}1-D_i^2-C_{ij}^2=C_{ik}^2\ge0.\tag{20}\label{eq:20}\end{equation}
Thus a decrease in local purity can be shared between two reduced pairs. Averaging Eq.~\eqref{eq:18} over all modes counts each pairwise squared concurrence twice and gives
\begin{equation}\FitColumnMath{D^2+\frac23\big(C_{e\mu}^2+C_{e\tau}^2+C_{\mu\tau}^2\big)=1.}\tag{21}\label{eq:21}\end{equation}

This global identity is consistent with the first-order-coherence--intrinsic-concurrence relations discussed in Refs.~\cite{ming2021,li2022}. Spin-flip-based connections between entanglement and mixedness provide the broader setting \cite{jaeger2003}; for the reduced state in Eq.~\eqref{eq:5}, the spin-flipped product has at most one nonzero eigenvalue, so intrinsic concurrence equals ordinary concurrence. Equation~\eqref{eq:21} is the corresponding single-excitation specialization. The local form in Eq.~\eqref{eq:18} resolves its allocation to a selected reference flavor. Figure~\ref{fig:2} illustrates the local balance for electron preparation; each panel resolves one reference mode into its local and two pairwise contributions.\label{discussion:e-coherence}

\subsection{Exact coherence--concurrence--CHSH relation}

\begin{corollary}\label{cor:pair}
For any reduced pair \(i,j\) of the pure three-mode single-excitation state, the optimized CHSH value is fixed by its average first-order coherence and concurrence through
\begin{equation}\FitColumnMath{\frac{B_{ij}^2}{4}=C_{ij}^2+\max\{C_{ij}^2,2D_{ij}^2+2C_{ij}^2-1\}.}\tag{22}\label{eq:22}\end{equation}
\end{corollary}

\begin{proof}
For an arbitrary two-qubit state, Hilbert--Schmidt orthogonality of the Pauli basis relates its purity to the local Bloch-vector lengths and the correlation tensor \cite{dutong2021}:
\begin{equation}\FitColumnMath{\operatorname{Tr}\rho_{ij}^2=\frac14\big(1+2D_{ij}^2+\operatorname{Tr}[T_{ij}^{\mathsf T}T_{ij}]\big).}\tag{23}\label{eq:23}\end{equation}
The complete three-mode state is pure, so the nonzero spectra of \(\rho_{ij}\) and \(\rho_k\) coincide. Their purities therefore obey
\begin{equation}\operatorname{Tr}\rho_{ij}^2=\operatorname{Tr}\rho_k^2=\frac{1+D_k^2}{2}.\tag{24}\label{eq:24}\end{equation}
Combining Eqs.~\eqref{eq:23} and~\eqref{eq:24} with the tensor spectrum in Eq.~\eqref{eq:16} yields
\begin{equation}\begin{aligned}
2C_{ij}^2+D_k^2&=1+2D_k^2-2D_{ij}^2,\\
D_k^2&=2D_{ij}^2+2C_{ij}^2-1.
\end{aligned}\tag{25}\label{eq:25}\end{equation}
Substitution of Eq.~\eqref{eq:25} into the optimized CHSH expression, Eq.~\eqref{eq:17}, gives Eq.~\eqref{eq:22}. This completes the proof.
\end{proof}

The proof uses purity of the complete state and does not assume that the selected two-mode reduction is pure. The third-mode population is retained: direct substitution of the flavor probabilities also gives \(D_{ij}^2=1-C_{ij}^2-2p_k(1-p_k)\). Consequently, \(D_{ij}\) and \(C_{ij}\) together determine the longitudinal tensor contribution, although neither quantity fixes it separately.

Introduce the ordering parameter \(\Gamma_{ij}=2D_{ij}^2+C_{ij}^2\). Equivalently,

\begin{equation}\frac{B_{ij}^2}{4}=\begin{cases}2C_{ij}^2,&\Gamma_{ij}\le1,\\2D_{ij}^2+3C_{ij}^2-1,&\Gamma_{ij}>1.\end{cases}\tag{26}\label{eq:26}\end{equation}
\begin{figure*}[t]\centering\includegraphics[width=.94\textwidth]{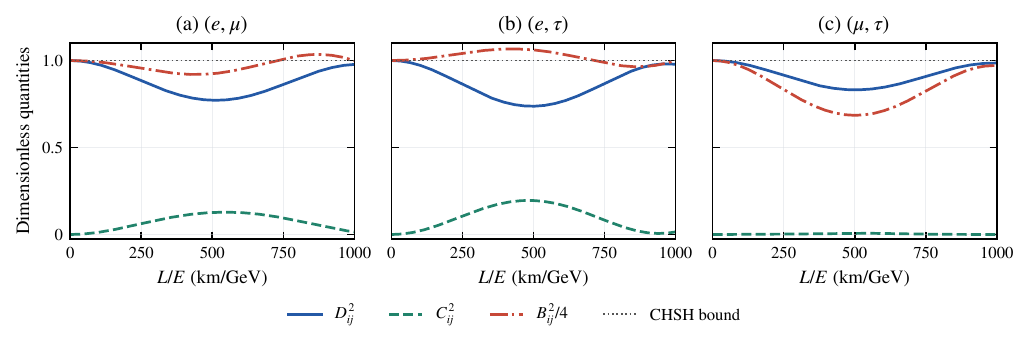}\caption{Average first-order coherence, concurrence, and optimized CHSH value for initially electron neutrinos. Panels (a)--(c) correspond to \((e,\mu)\), \((e,\tau)\), and \((\mu,\tau)\). Solid blue, dashed green, and dash-dotted red curves show \(D_{ij}^2\), \(C_{ij}^2\), and \(B_{ij}^2/4\) versus \(L/E\), related by Eq.~\eqref{eq:22}. The dotted horizontal line marks the CHSH bound; Bell violation occurs when the red curve exceeds unity.}\label{fig:3}\end{figure*}

\noindent This proves the exact relationship for every reduced pair of the three-mode pure state. In the first branch, the two transverse eigenvalues \(C_{ij}^2\) are the largest, so \(B_{ij}\) is proportional to concurrence. In the second branch, the longitudinal eigenvalue \(D_k^2\) exceeds one transverse eigenvalue and must enter the CHSH optimization. The branches agree at \(2D_{ij}^2+C_{ij}^2=1\). A change of slope is therefore an eigenvalue-ordering effect; it does not indicate discontinuous state evolution.\label{discussion:e-chsh}

The same equation can be represented by two planar pieces in squared variables:

\begin{equation}\FitColumnMath{\begin{aligned}
\frac{B_{ij}^2}{8}-C_{ij}^2&=0\quad(\Gamma_{ij}\le1),\\
2D_{ij}^2+3C_{ij}^2-\frac{B_{ij}^2}{4}&=1\quad(\Gamma_{ij}>1).
\end{aligned}}\tag{27}\label{eq:27}\end{equation}
\noindent Bell violation is characterized exactly by

\begin{equation}\FitColumnMath{B_{ij}>2\quad\Longleftrightarrow\quad C_{ij}^2>\tfrac12\;\text{or}\;2D_{ij}^2+3C_{ij}^2>2.}\tag{28}\label{eq:28}\end{equation}

These conditions are evaluated on physically allowed \((C_{ij}^2,D_{ij}^2)\) pairs. In particular, a nonzero concurrence need not meet either condition. For the equal-probability W state, \(C_{ij}^2=4/9\), \(D_{ij}^2=1/9\), and \(B_{ij}=4\sqrt2/3<2\) for every pair. In the two-flavor limit, \(D_{ij}^2=1-C_{ij}^2\), and Eq.~\eqref{eq:22} reduces to

\begingroup
\setlength{\belowdisplayskip}{0pt}
\setlength{\belowdisplayshortskip}{0pt}
\begin{equation}\begin{aligned}D_{ij}^2+C_{ij}^2&=1,\\B_{ij}^2&=4(1+C_{ij}^2),\\D_{ij}^2+\frac{B_{ij}^2}{4}&=2.\end{aligned}\tag{29}\label{eq:29}\end{equation}
\endgroup
\noindent Equation~\eqref{eq:29} is recovered only when the discarded mode is unoccupied; tracing out an occupied flavor does not give the same relation.

\section{Flavor evolution of first-order coherence, concurrence, and Bell correlations}
\begin{figure*}[t]\centering\includegraphics[width=\textwidth]{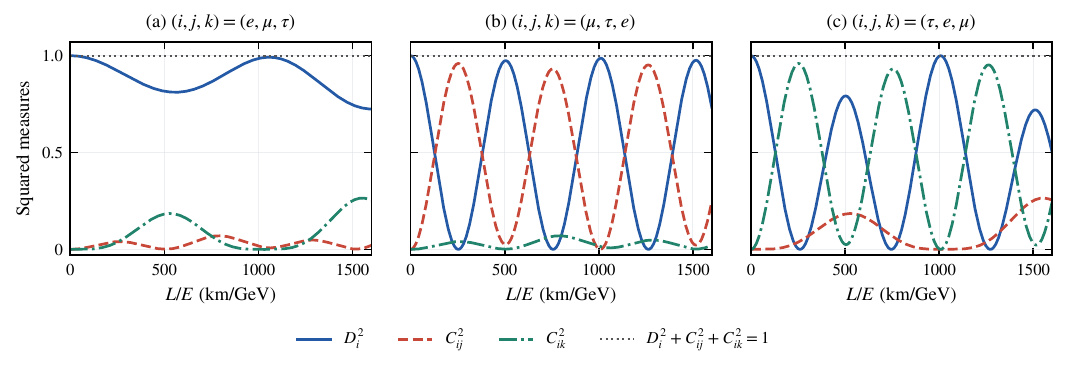}\caption{First-order coherence and pairwise concurrence for initially muon neutrinos. The three panels, cyclic mode orderings, and line conventions are those of Fig. 2, with \(0\le L/E\le1600\,\mathrm{km/GeV}\). The dotted sum remains unity for every reference mode. The curves use the full normalized three-flavor vacuum probabilities.}\label{fig:4}\end{figure*}

We apply the unified relations to electron and muon preparation. Figure~\ref{fig:1} compares the survival probabilities, Figs.~\ref{fig:2} and~\ref{fig:3} describe electron-initial correlations, and Figs.~\ref{fig:4} and~\ref{fig:5} describe muon-initial correlations.

\subsection{Initially electron neutrinos}
For \(\alpha=e\), the first row of the PMNS matrix gives
\begin{equation}\FitColumnMath{\begin{aligned}
P_{ee}={}&1-c_{13}^4\sin^2(2\theta_{12})\sin^2\Delta_{21}\\
&-\sin^2(2\theta_{13})\big(c_{12}^2\sin^2\Delta_{31}+s_{12}^2\sin^2\Delta_{32}\big).
\end{aligned}}\tag{30}\label{eq:30}\end{equation}
\noindent The atmospheric phases govern the shallow dip in Fig.~\ref{fig:1}(a), where \(P_{ee}\simeq0.9116\) near \(502.1\,\mathrm{km/GeV}\). The small value of \(\theta_{13}\) limits the disappearance amplitude, so the electron flavor retains most of the occupation throughout this interval. After the minimum, the survival probability increases toward a partial revival. The solar term and the difference between the two atmospheric phases keep this recovery below an exact return to the initial state within the displayed range.

The 33 Daya Bay central ratios from EH1, EH2, and EH3 illustrate the short-baseline disappearance pattern \cite{dayabay2015,ming2020}. The near-hall markers mainly sample the descending side, while the far-hall markers cover the dip and recovery. These reconstructed ratios accompany the unsmeared vacuum curve. The separate appearance probabilities used below follow from the three-flavor amplitudes, since \(P_{ee}\) fixes only \(p_e\) and the sum \(p_\mu+p_\tau\).

Writing \(p_i^{(e)}=P_{ei}\), the electron-mode first-order coherence and concurrence obey
\begin{equation}\begin{aligned}(D_e^{(e)})^2&=1-(C_{e\mu}^{(e)})^2-(C_{e\tau}^{(e)})^2,\\(C_{ei}^{(e)})^2&=4P_{ee}P_{ei}.\end{aligned}\tag{31}\label{eq:31}\end{equation}
\noindent Here the superscript denotes the initial flavor. At preparation, \(D_i^2=1\) and \(C_{ij}=0\). As electron survival decreases, the reduction of \(D_e^2\) is shared between the \(e\mu\) and \(e\tau\) squared concurrences in Fig.~\ref{fig:2}(a). The appearance-mode balances in panels (b),(c) retain high first-order coherence because the muon and tau populations remain small: a nearly empty mode also has a nearly pure one-mode reduction. The maximal concurrences are approximately \(C_{e\mu}=0.359\), \(C_{e\tau}=0.443\), and \(C_{\mu\tau}=0.086\). The weak \(\mu\tau\) contribution involves the product of two small appearance populations, whereas the electron-containing concurrences depend on one large survival population. Their unequal amplitudes reflect the partition of appearance probability between the two other flavors. The dotted sum stays at unity for every reference mode, showing that the decrease of first-order coherence is accounted for by both incident squared concurrences, as required by Eq.~\eqref{eq:18}.

% Reflow this paragraph by two lines to fill the page-6 right column.
\looseness=2
Figure~\ref{fig:3} displays the joint evolution of \(D_{ij}^2\), \(C_{ij}^2\), and \(B_{ij}^2/4\), with Bell violation indicated by the CHSH curve exceeding the dotted bound at unity. All three pairs satisfy \(\Gamma_{ij}>1\) over the displayed electron-initial interval and follow the second branch of Eq.~\eqref{eq:26}, where both first-order coherence and concurrence contribute to the optimized Bell value. The electron-containing pairs reach \(B_{e\mu}\simeq2.035\) near \(870.4\,\mathrm{km/GeV}\) and \(B_{e\tau}\simeq2.066\) near \(424.4\,\mathrm{km/GeV}\). Their CHSH curves cross the bound only slightly, while concurrence remains nonzero over broader intervals. The Bell maxima are displaced from the concurrence maxima because the third-mode population, and hence the longitudinal term in Eq.~\eqref{eq:17}, varies during propagation. The \(\mu\tau\) pair develops a small concurrence but does not exceed the CHSH bound. The three panels therefore distinguish the presence of pairwise entanglement from Bell violation and show why concurrence alone cannot determine the location or strength of the Bell maximum.

\subsection{Initially muon neutrinos}
\begin{figure*}[t]\centering\includegraphics[width=\textwidth]{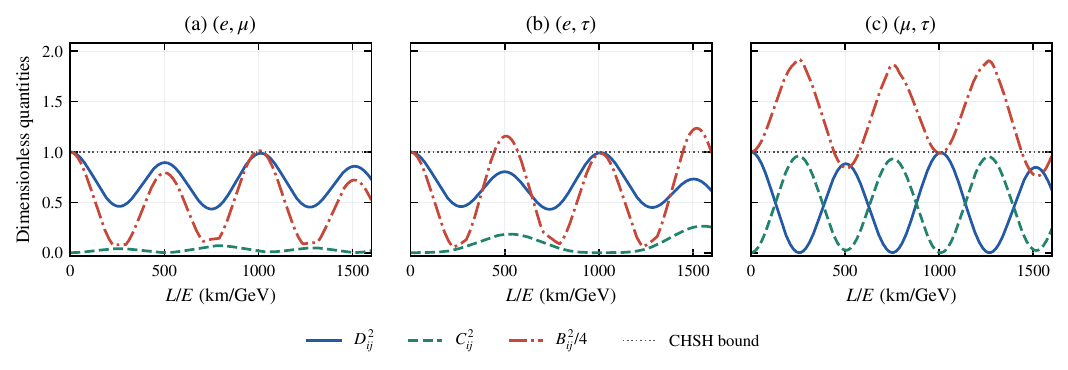}\caption{Average first-order coherence, concurrence, and optimized CHSH value for initially muon neutrinos. The panels and line conventions are those of Fig. 3, with \(0\le L/E\le1600\,\mathrm{km/GeV}\). The dotted line marks \(B_{ij}^2/4=1\). The curves are theoretical predictions; the dominant \(\mu\tau\) pair exhibits the strongest CHSH violation in this interval.}\label{fig:5}\end{figure*}

For \(\alpha=\mu\), Eq.~\eqref{eq:3} yields
\begin{equation}\FitColumnMath{\begin{aligned}
P_{\mu\mu}={}&1-4|U_{\mu1}|^2|U_{\mu2}|^2\sin^2\Delta_{21}\\
&-4|U_{\mu3}|^2\big(|U_{\mu1}|^2\sin^2\Delta_{31}+|U_{\mu2}|^2\sin^2\Delta_{32}\big).
\end{aligned}}\tag{32}\label{eq:32}\end{equation}
\noindent Near-maximal atmospheric mixing produces almost complete disappearance in Fig.~\ref{fig:1}(b): the first minimum is \(P_{\mu\mu}\simeq0.0065\) near \(505.4\,\mathrm{km/GeV}\). A subsequent revival and a second minimum near \(1516\,\mathrm{km/GeV}\) display the strong flavor conversion of muon preparation. The logarithmic abscissa resolves the low-\(L/E\) plateau and the later disappearance structure. The 39 reconstructed MINOS markers are rate ratios, including values above unity, and retain the observable convention of the source figure \cite{sousa2015,ming2020}. The normalized theoretical probabilities provide the state used to calculate the correlations in Figs.~\ref{fig:4} and~\ref{fig:5}.

For \(p_i^{(\mu)}=P_{\mu i}\), the muon-mode balance becomes
\begin{equation}\FitColumnMath{(D_\mu^{(\mu)})^2=1-(C_{\mu e}^{(\mu)})^2-(C_{\mu\tau}^{(\mu)})^2.}\tag{33}\label{eq:33}\end{equation}
\noindent The dominant exchange of occupation is between the muon and tau modes. Their concurrence reaches \(C_{\mu\tau}\simeq0.980\) near \(253.4\,\mathrm{km/GeV}\), where the two populations are comparable. This maximum precedes the first survival minimum: once the muon population becomes small, both incident concurrences are suppressed and \(D_\mu^2\) returns toward unity. The balances in Fig.~\ref{fig:4}(b),(c) consequently display large oscillations between first-order coherence and the \(\mu\tau\) contribution. This behavior follows from the dependence of concurrence on the population product, rather than on the amount of disappearance alone. The electron-mode balance in panel (a) varies more weakly, although electron appearance becomes more relevant at longer distances, with maximal \(C_{e\mu}\simeq0.266\) and \(C_{e\tau}\simeq0.515\). The dotted sum remains unity in every panel. Thus the dominant \(\mu\tau\) exchange approximates two-flavor sharing over part of the evolution, while the smaller electron-containing terms retain the full three-flavor balance.\label{discussion:mu-coherence}

In Fig.~\ref{fig:5}(c), the \(\mu\tau\) pair reaches \(B_{\mu\tau}\simeq2.771\) near \(253.4\,\mathrm{km/GeV}\). This maximum lies on the first branch of Eq.~\eqref{eq:26}, where \(B_{\mu\tau}=2\sqrt2\,C_{\mu\tau}\), so the concurrence and CHSH maxima coincide. The proportionality arises because the two transverse concurrence eigenvalues determine the Bell optimization at that point. During the full muon-initial evolution, each pair samples both branches, producing changes in the relative shapes of the first-order coherence, concurrence, and Bell curves. The remaining pairs have distinct later violations: \(B_{e\mu}\simeq2.013\) near \(1004.8\,\mathrm{km/GeV}\) and \(B_{e\tau}\simeq2.223\) near \(1520.2\,\mathrm{km/GeV}\). The larger late \(e\tau\) violation develops as the discarded muon population becomes small, reducing the vacuum contribution to that pair in Eq.~\eqref{eq:5}. These changes reflect the competition between transverse concurrence and longitudinal third-mode terms in Eq.~\eqref{eq:17}. All pair trajectories nevertheless obey the same closed relation, Eq.~\eqref{eq:22}.\label{discussion:mu-chsh}

\section{Conclusions}
In this paper, we have established exact relations among first-order coherence, pairwise concurrence, and the optimized CHSH value for coherent three-flavor neutrino oscillations. The three-qubit single-excitation representation gives generally mixed two-qubit reductions and the balance \(D_i^2+C_{ij}^2+C_{ik}^2=1\), which resolves the contribution of each flavor pair. Eliminating the third-flavor population yields the closed relation in Eq.~\eqref{eq:22}; its two branches identify the correlation-tensor components determining the Bell value and provide exact conditions for CHSH violation. For initially electron neutrinos, the electron-containing pairs exhibit weak violation and displaced concurrence and CHSH maxima, whereas muon preparation produces a dominant \(\mu\tau\) pair whose maxima can coincide on the transverse branch. These results show how the third flavor changes the two-flavor correspondence and why maximal disappearance does not imply maximal entanglement. The disappearance-data overlays illustrate the oscillation pattern, while the correlation predictions require the complete three-flavor state. Extensions to coherent matter propagation and wave-packet-induced mixed states can further clarify how these relations depend on the physical propagation conditions.

\begin{acknowledgements}
\justifying\sloppy
\setlength{\emergencystretch}{2em}
This work was supported by the Anhui Provincial University Scientific Research Key Project (Grant No.\ 2025AHGXZK30635), the Natural Science Foundation of Anhui Sanlian University (Grant Nos.\ KJZD2025021 and KJZD2025020), the National Science Foundation of China (Grant No.\ 12205047), the Key Project for Cultivating Outstanding Young Teachers in the Training Program for Middle-aged and Young Teachers in Anhui Province (Grant No.\ YQZD2025094), the Natural Science Research Project of Education Department of Anhui Province of China (Grant No.\ 2022AH040199), the Anhui Provincial Natural Science Foundation (Grant No.\ 2308085QF195), and the Talent Funds of Hefei Institute of Technology (Grant Nos.\ 2025KY04, 2025KY05 and 2025KY07).
\end{acknowledgements}

\par\medskip
\noindent\textbf{Data Availability Statement}
The fixed oscillation inputs, theoretical curve tables, and reconstructed experimental central-point tables are included in the accompanying project. The immediate source figures and extraction provenance are retained. The five numbered figures use fixed central inputs throughout.

\par\medskip
\noindent\textbf{Code Availability Statement}
The calculation and plotting script and numerical verification report are included in the accompanying project. The correlation formulas have been checked against direct density-matrix concurrence and correlation-tensor eigenvalues.


\begin{thebibliography}{99}
\bibitem{maki1962} Z. Maki, M. Nakagawa, S. Sakata, \ReferenceLink{https://doi.org/10.1143/ptp.28.870}{\emph{Prog. Theor. Phys.} \textbf{28}, 870--880 (1962)}.

\bibitem{giuntikimbook} C. Giunti, C. W. Kim, \ReferenceLink{https://doi.org/10.1093/acprof:oso/9780198508717.001.0001}{\emph{Fundamentals of Neutrino Physics and Astrophysics} (Oxford University Press, Oxford, 2007)}.

\bibitem{akhmedov2009} E. K. Akhmedov, A. Y. Smirnov, \ReferenceLink{https://doi.org/10.1134/s1063778809080122}{\emph{Phys. At. Nucl.} \textbf{72}, 1363--1381 (2009)}.

\bibitem{superk1998} Y. Fukuda, T. Hayakawa, E. Ichihara et al., \ReferenceLink{https://doi.org/10.1103/physrevlett.81.1562}{\emph{Phys. Rev. Lett.} \textbf{81}, 1562--1567 (1998)}.

\bibitem{sno2002} Q. R. Ahmad, R. C. Allen, T. C. Andersen et al., \ReferenceLink{https://doi.org/10.1103/physrevlett.89.011301}{\emph{Phys. Rev. Lett.} \textbf{89}, 011301 (2002)}.

\bibitem{dayabay2012} F. P. An, J. Z. Bai, A. B. Balantekin et al., \ReferenceLink{https://doi.org/10.1103/physrevlett.108.171803}{\emph{Phys. Rev. Lett.} \textbf{108}, 171803 (2012)}.

\bibitem{minos2008} P. Adamson, C. Andreopoulos, K. E. Arms et al., \ReferenceLink{https://doi.org/10.1103/physrevlett.101.131802}{\emph{Phys. Rev. Lett.} \textbf{101}, 131802 (2008)}.

\bibitem{minos2014} P. Adamson, I. Anghel, A. Aurisano et al., \ReferenceLink{https://doi.org/10.1103/physrevlett.112.191801}{\emph{Phys. Rev. Lett.} \textbf{112}, 191801 (2014)}.

\bibitem{dayabay2015} F. P. An, A. B. Balantekin, H. R. Band et al., \ReferenceLink{https://doi.org/10.1103/physrevlett.115.111802}{\emph{Phys. Rev. Lett.} \textbf{115}, 111802 (2015)}.

\bibitem{blasone2008mixing} M. Blasone, F. Dell'Anno, S. De Siena, M. Di Mauro, F. Illuminati, \ReferenceLink{https://doi.org/10.1103/physrevd.77.096002}{\emph{Phys. Rev. D} \textbf{77}, 096002 (2008)}.

\bibitem{blasone2008} M. Blasone, F. Dell'Anno, S. De Siena, F. Illuminati, \ReferenceLink{https://doi.org/10.1209/0295-5075/85/50002}{\emph{EPL} \textbf{85}, 50002 (2009)}.

\bibitem{banerjee2015three} S. Banerjee, A. K. Alok, R. Srikanth, B. C. Hiesmayr, \ReferenceLink{https://doi.org/10.1140/epjc/s10052-015-3717-x}{\emph{Eur. Phys. J. C} \textbf{75}, 487 (2015)}.

\bibitem{banerjee2015} A. K. Alok, S. Banerjee, S. Uma Sankar, \ReferenceLink{https://doi.org/10.1016/j.nuclphysb.2016.05.001}{\emph{Nucl. Phys. B} \textbf{909}, 65--72 (2016)}.

\bibitem{song2018} X.-K. Song, Y. Huang, J. Ling, M.-H. Yung, \ReferenceLink{https://doi.org/10.1103/physreva.98.050302}{\emph{Phys. Rev. A} \textbf{98}, 050302 (2018)}.

\bibitem{formaggio2016} J. A. Formaggio, D. I. Kaiser, M. M. Murskyj, T. E. Weiss, \ReferenceLink{https://doi.org/10.1103/physrevlett.117.050402}{\emph{Phys. Rev. Lett.} \textbf{117}, 050402 (2016)}.

\bibitem{ming2020} F. Ming, X.-K. Song, J. Ling, L. Ye, D. Wang, \ReferenceLink{https://doi.org/10.1140/epjc/s10052-020-7840-y}{\emph{Eur. Phys. J. C} \textbf{80}, 275 (2020)}.

\bibitem{li2022} Y.-W. Li, L.-J. Li, X.-K. Song, D. Wang, \ReferenceLink{https://doi.org/10.1140/epjp/s13360-022-03459-z}{\emph{Eur. Phys. J. Plus} \textbf{137}, 1267 (2022)}.

\bibitem{bittencourt2022} V. A. S. V. Bittencourt, M. Blasone, S. De Siena, C. Matrella, \ReferenceLink{https://doi.org/10.1140/epjc/s10052-022-10508-5}{\emph{Eur. Phys. J. C} \textbf{82}, 566 (2022)}.

\bibitem{bittencourt2024} V. A. S. V. Bittencourt, M. Blasone, S. De Siena, C. Matrella, \ReferenceLink{https://doi.org/10.1140/epjc/s10052-024-12631-x}{\emph{Eur. Phys. J. C} \textbf{84}, 301 (2024)}.

\bibitem{blasone2021} M. Blasone, S. De Siena, C. Matrella, \ReferenceLink{https://doi.org/10.1140/epjc/s10052-021-09471-4}{\emph{Eur. Phys. J. C} \textbf{81}, 660 (2021)}.

\bibitem{dayabay2017} Daya Bay Collaboration, \ReferenceLink{https://doi.org/10.1140/epjc/s10052-017-4970-y}{\emph{Eur. Phys. J. C} \textbf{77}, 606 (2017)}.

\bibitem{dong2022} D.-D. Dong, G.-B. Wei, X.-K. Song, D. Wang, L. Ye, \ReferenceLink{https://doi.org/10.1103/physreva.106.042415}{\emph{Phys. Rev. A} \textbf{106}, 042415 (2022)}.

\bibitem{dong2023} D.-D. Dong, X.-K. Song, X.-G. Fan, L. Ye, D. Wang, \ReferenceLink{https://doi.org/10.1103/physreva.107.052403}{\emph{Phys. Rev. A} \textbf{107}, 052403 (2023)}.

\bibitem{chitambar2019} E. Chitambar, G. Gour, \ReferenceLink{https://doi.org/10.1103/revmodphys.91.025001}{\emph{Rev. Mod. Phys.} \textbf{91}, 025001 (2019)}.

\bibitem{horodecki2009} R. Horodecki, P. Horodecki, M. Horodecki, K. Horodecki, \ReferenceLink{https://doi.org/10.1103/revmodphys.81.865}{\emph{Rev. Mod. Phys.} \textbf{81}, 865--942 (2009)}.

\bibitem{streltsov2017} A. Streltsov, G. Adesso, M. B. Plenio, \ReferenceLink{https://doi.org/10.1103/revmodphys.89.041003}{\emph{Rev. Mod. Phys.} \textbf{89}, 041003 (2017)}.

\bibitem{baumgratz2014} T. Baumgratz, M. Cramer, M. B. Plenio, \ReferenceLink{https://doi.org/10.1103/physrevlett.113.140401}{\emph{Phys. Rev. Lett.} \textbf{113}, 140401 (2014)}.

\bibitem{svozilik2015} J. Svozilik, A. Valles, J. Perina, J. P. Torres, \ReferenceLink{https://doi.org/10.1103/physrevlett.115.220501}{\emph{Phys. Rev. Lett.} \textbf{115}, 220501 (2015)}.

\bibitem{chsh1969} J. F. Clauser, M. A. Horne, A. Shimony, R. A. Holt, \ReferenceLink{https://doi.org/10.1103/physrevlett.23.880}{\emph{Phys. Rev. Lett.} \textbf{23}, 880--884 (1969)}.

\bibitem{horodecki1995} R. Horodecki, P. Horodecki, M. Horodecki, \ReferenceLink{https://doi.org/10.1016/0375-9601(95)00214-n}{\emph{Phys. Lett. A} \textbf{200}, 340--344 (1995)}.

\bibitem{wiseman2007} H. M. Wiseman, S. J. Jones, A. C. Doherty, \ReferenceLink{https://doi.org/10.1103/physrevlett.98.140402}{\emph{Phys. Rev. Lett.} \textbf{98}, 140402 (2007)}.

\bibitem{uola2020} R. Uola, A. C. S. Costa, H. C. Nguyen, O. Guhne, \ReferenceLink{https://doi.org/10.1103/revmodphys.92.015001}{\emph{Rev. Mod. Phys.} \textbf{92}, 015001 (2020)}.

\bibitem{cavalcanti2009} E. G. Cavalcanti, S. J. Jones, H. M. Wiseman, M. D. Reid, \ReferenceLink{https://doi.org/10.1103/physreva.80.032112}{\emph{Phys. Rev. A} \textbf{80}, 032112 (2009)}.

\bibitem{costa2016} A. C. S. Costa, R. M. Angelo, \ReferenceLink{https://doi.org/10.1103/physreva.93.020103}{\emph{Phys. Rev. A} \textbf{93}, 020103 (2016)}.

\bibitem{fan2022} Y. Fan, C. Jia, L. Qiu, \ReferenceLink{https://doi.org/10.1103/physreva.106.012433}{\emph{Phys. Rev. A} \textbf{106}, 012433 (2022)}.

\bibitem{englert1996} B.-G. Englert, \ReferenceLink{https://doi.org/10.1103/physrevlett.77.2154}{\emph{Phys. Rev. Lett.} \textbf{77}, 2154--2157 (1996)}.

\bibitem{jakob2010} M. Jakob, J. A. Bergou, \ReferenceLink{https://doi.org/10.1016/j.optcom.2009.10.044}{\emph{Opt. Commun.} \textbf{283}, 827--830 (2010)}.

\bibitem{basso2020} M. L. W. Basso, J. Maziero, \ReferenceLink{https://doi.org/10.1088/1751-8121/abc361}{\emph{J. Phys. A: Math. Theor.} \textbf{53}, 465301 (2020)}.

\bibitem{bassomixed2021} M. L. W. Basso, J. Maziero, \ReferenceLink{https://doi.org/10.1209/0295-5075/ac1bc8}{\emph{EPL} \textbf{135}, 60002 (2021)}.

\bibitem{dutong2021} M.-M. Du, D. M. Tong, \ReferenceLink{https://doi.org/10.1103/physreva.103.032407}{\emph{Phys. Rev. A} \textbf{103}, 032407 (2021)}.

\bibitem{ming2021} F. Ming, D. Wang, L.-J. Li et al., \ReferenceLink{https://doi.org/10.1002/qute.202100036}{\emph{Adv. Quantum Technol.} \textbf{4}, 2100036 (2021)}.

\bibitem{gravity2024} D.-D. Chen, D.-D. Dong, X.-K. Song, L. Ye, D. Wang, \ReferenceLink{https://doi.org/10.1140/epjc/s10052-024-12616-w}{\emph{Eur. Phys. J. C} \textbf{84}, 254 (2024)}.

\bibitem{kayser1981} B. Kayser, \ReferenceLink{https://doi.org/10.1103/physrevd.24.110}{\emph{Phys. Rev. D} \textbf{24}, 110--116 (1981)}.

\bibitem{giuntikim1998} C. Giunti, C. W. Kim, \ReferenceLink{https://doi.org/10.1103/physrevd.58.017301}{\emph{Phys. Rev. D} \textbf{58}, 017301 (1998)}.

\bibitem{giunti2004} C. Giunti, \ReferenceLink{https://doi.org/10.1023/b:fopl.0000019651.53280.31}{\emph{Found. Phys. Lett.} \textbf{17}, 103--124 (2004)}.

\bibitem{kiers1996} K. Kiers, S. Nussinov, N. Weiss, \ReferenceLink{https://doi.org/10.1103/physrevd.53.537}{\emph{Phys. Rev. D} \textbf{53}, 537--547 (1996)}.

\bibitem{stodolsky1998} L. Stodolsky, \ReferenceLink{https://doi.org/10.1103/physrevd.58.036006}{\emph{Phys. Rev. D} \textbf{58}, 036006 (1998)}.

\bibitem{blasone1995} M. Blasone, G. Vitiello, \ReferenceLink{https://doi.org/10.1006/aphy.1995.1115}{\emph{Ann. Phys.} \textbf{244}, 283--311 (1995)}.

\bibitem{beuthe2003} M. Beuthe, \ReferenceLink{https://doi.org/10.1016/s0370-1573(02)00538-0}{\emph{Phys. Rep.} \textbf{375}, 105--218 (2003)}.

\bibitem{akhmedovkopp2010} E. K. Akhmedov, J. Kopp, \ReferenceLink{https://doi.org/10.1007/jhep04(2010)008}{\emph{JHEP} \textbf{04}, 008 (2010)}.

\bibitem{nufit2024} I. Esteban, M. C. Gonzalez-Garcia, M. Maltoni, I. Martinez-Soler, J. P. Pinheiro, T. Schwetz, \ReferenceLink{https://doi.org/10.1007/jhep12(2024)216}{\emph{JHEP} \textbf{12}, 216 (2024)}.

\bibitem{sousa2015} A. B. Sousa (for the MINOS and MINOS+ Collaborations), \ReferenceLink{https://doi.org/10.1063/1.4915576}{\emph{AIP Conf. Proc.} \textbf{1666}, 110004 (2015)}.

\bibitem{kamland2013} A. Gando et al. (KamLAND Collaboration), \ReferenceLink{https://doi.org/10.1103/PhysRevD.88.033001}{\emph{Phys. Rev. D} \textbf{88}, 033001 (2013)}.

\bibitem{minos2011} P. Adamson et al. (MINOS Collaboration), \ReferenceLink{https://doi.org/10.1103/PhysRevLett.106.181801}{\emph{Phys. Rev. Lett.} \textbf{106}, 181801 (2011)}.

\bibitem{t2k2017} K. Abe et al. (T2K Collaboration), \ReferenceLink{https://doi.org/10.1103/PhysRevD.96.011102}{\emph{Phys. Rev. D} \textbf{96}, 011102 (2017)}.

\bibitem{streltsov2015} A. Streltsov, U. Singh, H. S. Dhar, M. N. Bera, G. Adesso, \ReferenceLink{https://doi.org/10.1103/physrevlett.115.020403}{\emph{Phys. Rev. Lett.} \textbf{115}, 020403 (2015)}.

\bibitem{mondal2017} D. Mondal, T. Pramanik, A. K. Pati, \ReferenceLink{https://doi.org/10.1103/physreva.95.010301}{\emph{Phys. Rev. A} \textbf{95}, 010301 (2017)}.

\bibitem{wootters1998} W. K. Wootters, \ReferenceLink{https://doi.org/10.1103/physrevlett.80.2245}{\emph{Phys. Rev. Lett.} \textbf{80}, 2245--2248 (1998)}.

\bibitem{coffman2000} V. Coffman, J. Kundu, W. K. Wootters, \ReferenceLink{https://doi.org/10.1103/physreva.61.052306}{\emph{Phys. Rev. A} \textbf{61}, 052306 (2000)}.

\bibitem{jaeger2003} G. Jaeger, A. V. Sergienko, B. E. A. Saleh, M. C. Teich, \ReferenceLink{https://doi.org/10.1103/physreva.68.022318}{\emph{Phys. Rev. A} \textbf{68}, 022318 (2003)}.
\end{thebibliography}
\end{document}